# The Ability of Virtual Reality Technologies to Improve Comprehension of Speech Therapy Device Training

Daniel E. Killough

Academic Magnet High School

December 12, 2018

## List of Abbreviations

| | |
|---|---|
| 2D | Two-Dimensional |
| 3D | Three-Dimensional |
| FOV | Field of view |
| OPT | Oral Placement Therapy |
| VE | Virtual Environment |
| VR | Virtual Reality |

## Abstract

This study evaluates the usage of virtual reality (VR) technologies as a teaching tool in oral placement therapy, a subset of speech therapy. The researcher distributed instructional videos using traditional lecture and modified three-dimensional video to prompt responses. Data was gathered with a two-part Google Form: In "Section 1: Knowledge Test" participants were asked to determine how well they received the information displayed to them. In "Section 2: Opinion Test" participants were asked diagnostic and subjective questions via Likert scale ranging from 1 ("Strongly Disagree") to 5 ("Strongly Agree") to determine how well they enjoyed viewing the information displayed to them. Averages for Section 1 were 92.00% for the control group (viewing 2D, unmodified video) and 77.88% for the experimental group (viewing 3D, VR video). Almost all participants answered at least 60% of the questions correctly. Averages for 2D and 3D participants were 4.53/5 and 3.82/5, respectively for "positive" prompts. Exactly 50% of participants experiencing VR video preferred the method to a traditional lecture. This study determines that virtual reality is viable as a learning tool, but knowledge obtained is not necessarily as high as using traditional lecture. Further experimentation is required to determine how well oral placement therapists respond to physically interacting with a model instead of only viewing it. Copies of the Google Form used to collect responses, all raw data, and a flowchart outlining each step used to construct the 3D video can be found in the Appendix.

# The Ability of Virtual Reality Technologies to Improve Comprehension of Speech Therapy Device Training

The purpose of this chapter is to provide an overview of how virtual reality technologies could be utilized to improve usage of speech therapy devices by oral placement therapists. Outlines of the research problem, the researcher's methods to conduct inquiries, and the project's significance are also included.

## The Topic

From spinal surgery to sporting events, from traveling the world to video gaming, virtual reality (VR) technologies have become increasingly commonplace over the past 20 years. Companies such as Sony, HTC, and Oculus have created high-resolution systems that completely encase a user's senses of sight and sound, transporting them into digital realms with their Playstation VR, Vive, and Rift platforms, respectively (Alaraj et al., 2013). These creative platforms allow developers to construct interactive data sets in three-dimensions (3D), as opposed to two-dimensional (2D) information on a flat computer monitor. Similar displays, such as Google Cardboard and Samsung Gear VR, utilize stereoscopic imaging, a process by which a single two-dimensional scene is divided into two "images [shown to] each eye, [generating the illusion of] a single object of three dimensions" using a smartphone (Virtual Reality Society, 2016, para. 3). Since much of virtual reality content is used to entertain, very few consider using VR as a realistic medium to convey factual information. Due to a common misconception that new, technologically-advanced products are solely used for leisure, it is necessary that interactive hardware be more widely adapted to broaden the scope of what is acceptable in a classroom environment while giving pupils a higher degree of comprehension of material being presented to them.

## The Research Problem

Completed studies such as Liarokapis et al.'s (2004) experiment with 3D Web content and Shelton and Hedley's (2002) instructional demonstration with an in-depth look into the solar system suggest a higher grasp of comprehension when interactive virtual multimedia content is present as compared to traditional, projected diagrams and text. Addressing concerns of the technology itself being too difficult to use or learn, Kaufmann (2003) highlights the ease of interacting with virtual objects by using "natural means of communication [such as] speech [or] gestures" (p. 1).

Although virtual reality platforms have the potential to revolutionize how people interact with one another, the technology is underutilized. Though some applications exist to take advantage of the hardware's capabilities, such as Google Tilt Brush and Google Earth VR (HTC Corporation, 2018), Zyda (2005) notes that "much of the research and development being conducted in the [VR community] parallels that of the [video gaming] community" (p. 25), ultimately allowing the "greater audience" (p. 25) to assume VR is ultimately an entertainment platform with no alternate uses. There is a disparity between the adoption of innovative technologies and the societal pretenses that precede them that ultimately must be closed in order to match the advancement of hardware.

## Purpose of the Study

The purpose of this study is to explore alternate uses of virtual reality hardware, specifically in the educational field, while enabling oral placement therapists to more effectively aid their patients. As a whole, the study focuses on technological innovations within the primary field of computer science with a secondary field of education to apply the computational commodity to a specific area of use. One desired outcome was similar to Liarokapis et al.'s study in 2004, albeit with more modern hardware, where the researchers found an increase in comprehension of the operation of mechanical systems after viewing digitally simulated three-dimensional web content; however, the researcher desired to immerse the user in a virtual environment in which the user may view three-dimensional muscular structures instead, demonstrating interactions between tools and muscular movements throughout the mouth. Strengths of learning via a virtual environment, as described by Rizzo and Kim (2005), include the ability to perform in a regulated, "safe testing and training environment" (p. 121) to allow the user to feel immersed in a scene without fear of outside influence. The governing question of the experiment is: How can virtual reality technologies be utilized in order to optimally train oral placement therapists and improve the comprehension, usage, and effectiveness of speech therapy tools? This study additionally answers sub-questions including:

- How have virtual reality technologies changed over their lifetime?
- What is the role of virtual environments within painting digital interactive landscapes?
- How does animated three-dimensional modeling effect visualizations of content, specifically with muscular system structure?

The significance of this study is to find a correlation between the effectiveness of virtual reality technologies and educational training experience to ultimately broaden its adoption across a wider range of fields and further the present to the future.

## Methods and Evaluation

This study answers the question of, “How can virtual reality technologies be utilized in order to optimally train oral placement therapists and improve the comprehension, usage, and effectiveness of speech therapy tools?” by distributing preexisting and modified video to the clients of TalkTools, a local oral placement therapy device company. As supported by Ware and Franck (1994), the amount of “information [that] can be perceived in [a 3D diagram is much greater than that of] 2D” (p. 182) from a “localized virtual reality display” (p. 182), demonstrating that in order to maximize results, 3D models should be used to compare with the traditional 2D format. More specifically, the methodology began by downloading a copy of a prerecorded lecture provided by TalkTools[1] and splitting the video into stereoscopic 3D. The researcher then recorded themselves using the 3D Organon VR Anatomy application in an Oculus Rift, which was then combined with the original lecture using Final Cut Pro X to give participants a realistic view inside the body of a model as opposed to imagining structures projected by an instructor. Participants were selected from TalkTools clients and therapists that owned or had access to a smartphone, and they were randomly assigned to control and experimental groups. Some clients were provided with the researcher-modified video and a free plastic VR viewer powered by a smartphone from Amazon, while others were only provided with a copy of the original, unmodified video. Note that the audio and material covered in both videos were identical. Data was gathered quantitatively and qualitatively via a Google Form made by the researcher, asking participants to recall key information from the videos and give their opinions on the information presented. Participants were given 5 points for every question they correctly answered on the knowledge test section and were scored out of 25 possible points. The opinion test section collected quantitative measures using a 5-point Likert scale questioning their engagement with the presentation. This study deductively compares the results of virtual reality video versus two-dimensional video, inductively observes how well therapists respond to each type, and creatively produces videos with an alternative, immersive educational approach. Success of this study was determined by observed participant retention of information as well as their reception to the video in order to judge the applicability of virtual reality technologies to oral placement therapy training and to what extent the results can be applied to a greater context. The conclusion was valid as since error was effectively minimized by distributing identical virtual reality viewers, along with attempting to eliminate consumer bias by selecting from a randomized group of therapists with the same or a very similar level of expertise. Additionally, variables pertaining to personal comfort such with motion sickness using the headsets are addressed in the survey to account for opinions pertaining to frequently expressed issues with the hardware. Results were analyzed by comparing form responses from participants in the control and experimental groups and determining which demonstrated greater understanding of the lecture. The answer to the governing question was derived from analysis results, demonstrating a proof of concept that VR can, objectively, be realistically expanded to the field of oral placement therapy, although further experimentation is necessary regarding the subject.

---

[1] Example lecture demonstration videos can be found on TalkTools’ website at https://talktools.com/pages/resources.

## Implications

Results of the study demonstrate a proof of concept as to whether or not virtual reality technologies are more or less viable for TalkTools speech therapy device training. Further implications not demonstrated directly from participant responses will depend on further research; however, the researcher determined that participants received the material being presented to them well in both virtual and physical environments. Consequences of the study may lead to users simply holding a preference to one learning method or another, perhaps biasing users to tend towards products that additionally offer 3D learning tools. Connections derived from these results include an expansion of the technology into other areas of study within the general field of education; furthermore, results of the study encourage the researcher to branch out experimentation with the Oculus Rift and similar hardware to program a full-length application in Unity3D, complete with head tracking and spacial awareness for all participants to fully immerse themselves in the material as well as have access to professional-quality headsets.

## Significance

The significance of this study relates to the field of computer science, which constantly evolves and applies to nearly every 21st century consumer. Many companies and device makers are constantly seeking to find innovative ways to further their understanding of the world around them. This study's results are able to bring more awareness to innovate hardware and may help spur development of applications that can be ultimately utilized to teach others about their careers.

## Applicability

The results of this thesis broaden the reach of virtual reality technologies, bringing more awareness to the hardware and its uses, specifically within education. Further research will further the application pool, allowing more software to be developed and further resources to be invested in improving the hardware. As supply and technology increase as companies release more hardware and ultimately lower its overall cost, device prices will perceivably become more reasonable for general consumers. Skills achieved from working on this project that will continue to be relevant include personally improving expertise with virtual reality hardware and video editing software as well as improving professional problem-solving and communication skills though conversations with mentors and advisors.

## A Review of the Literature Pertaining to Virtual Reality Technologies and Speech Therapy Device Training

Although literature concerning virtual reality technologies predominantly pertains to entertainment and video gaming, this study attempts to bring light to the applications and usage of currently released hardware and software within the realm of educational instruction. This review primarily focuses on individualized instructor training as well as oral muscular therapy within the sphere of preparatory learning while identifying and acknowledging innovations throughout the history of virtual reality technology. Sub-topics of the information discussed in this review include how VR technologies can best be utilized to fit a specific application; specific innovations within education, educational medicine, rehabilitation therapy, and physical therapy; virtual environments and their role with painting digital interactive landscapes; three-dimensional modeling, specifically with muscular system structure; and specific applications of tools themselves within the field of oral speech therapy. This review of literature ultimately strives to find the solution to the question: How can virtual reality technologies be best utilized in order to train oral speech therapists and improve the comprehension, usage, and effectiveness of speech therapy tools? by identifying and synthesizing information from previous studies as well as experimenting with head-mounted displays [HMD] such as the Oculus Rift to create and modify three-dimensional models of anatomically correct muscular systems.

### Innovations in Virtual Reality Technologies

From the mid-nineteenth to the early twenty-first century, research and development within virtual reality technologies has greatly modified the appearance, effectiveness, and usage of the hardware as a whole. Large, scenic murals such as Bartholomeus van der Helst’s 1648 oil painting *Banquet at the Crossbowmen's Guild in Celebration of the Treaty of Münster,* a detailed 2.3m x 5.5m oil on canvas piece depicting the signing of the Treaty of Münster, and Oscar E. Berninghaus' (n.d.) *Indian Attack on the Village of Saint Louis, 1780* depicting the British and Indian assault on St. Louis, Missouri during the Revolutionary War, are some of the first recorded incidences of objects created in order to make observers believe they are somewhere that they are not. However, virtual reality as known today did not arise until the research of Charles Wheatstone in 1838 "demonstrated that the brain processes [slightly shifted] two-dimensional images [shown to] each eye [as] a single object of three dimensions," (Virtual Reality Society, 2016, para. 3) ultimately producing the stereoscope: the first modern instance of using two split images to create a sense of depth. Today, stereoscopic technology continues to be "used [by] low-budget . . . head mounted displays for mobile phones" (para. 3), such as the Google Cardboard and Samsung Gear VR, by splitting a two-dimensional image into two slightly altered parts, projecting a false sense of depth and simulated realism to the observer (Thompson, 2017).

More elaborate sensory machines did not arise until the mid-1950's with the invention of "cinematographer Morton [Heilig's] Sensorama" (Virtual Reality Society, 2016, para. 6) which combined Wheatstone's stereoscopic display with stereo speakers, "fans, smell generators, and a vibrating chair, [which was] intended to fully immerse the individual in [Heilig's] film" (para. 6). Heilig's invention of the Telesphere Mask soon after in 1960 also marked "the first example of a head mounted display [that featured] stereoscopic [video] with stereo sound" (para. 7) set a precedent for later technologies. Charles Comeau and James Bryan's Headsight, which arose the following year, included integrated motion tracking systems embedded into the head-mounted display, allowing users to “naturally look around the environment” (Virtual Reality Society, 2016, para. 8) being displayed in front of them, though it lacked the ability to generate its own images without use of an external camera (Steinicke, 2016). However, in 1965 Ivan Sutherland created the concept of the “Ultimate Display” which “could simulate reality to the point where one could not tell the difference from actual reality” (Virtual Reality Society, 2016, para. 9) using fully immersive “computer-generated environments via novel types of multimodal input and output devices” (Steinicke, 2016, p. 19), an idealistic creation that quickly “[became] a core blueprint for the concepts that encompass virtual reality today” (Virtual Reality Society, 2016, para. 12). “Virtual Reality”, the name itself, was not known publicly until 1987, when Jaron Lanier popularized the term by detailing his research in his company’s visual programming lab. Lanier and his team’s developments later contributed to the investment and use of haptics in conjunction with head mounted displays by their commercialization of goggles and gloves (Virtual Reality Society, 2016, para. 15).

Since the turn of the century, rapidly expanding technological capabilities and investment in new products has enabled new products to enter the market at lower cost than their older, less efficient counterparts. Existing companies such as Sony and HTC now rival newly born agencies such as Oculus (acquired by Facebook in 2014 (Dredge, 2014)) to develop new, innovative head-mounted displays with complex architecture. Today, headsets come complete with sensors that use infrared positional tracking to acutely determine one’s location within a given space, allowing applications to react to one’s physical body movements and gestures as opposed to simply their head movements. Some models, such as the current flagship headsets the HTC Vive and Oculus Rift, include handheld controllers and/or headphones (Nield, 2016, para. 12; HTC

Corporation, 2018; Oculus VR, LLC, 2018b). The Oculus Rift's Touch controllers also ergonomically enable users to point, grab, and punch at objects in VR space, their movements transferred from their body to their hardware. Inside each headset resides a magnetometer, a gyroscope, and an accelerometer that, combined with infrared tracking, precisely triangulate the relative location of the headset in a given space. Using this information, the display determines the position of the user and, combined with other cameras on the headset and outside the user's field of reference, adjusts the location of objects on-screen and effects in-ear accordingly (Nield, 2016, para. 17).

Increased creation and adoption of VR hardware and software only enlarges the pool of developer resources and creativity, heightening the quantity and overall quality of simulations only possible within virtual environments. As contrasted in a study by Carrozzino and Bergamasco (2010), the use of desktop devices such as a monitor or desktop speakers as compared to wearables such as a head-mounted display or haptic engines show that, when categorized, the former was determined to be completely "Non-immersive" while demonstrating the latter as having "High Immersion" in some instances (p. 454). Companies that have adopted the use and manufacture of virtual reality technologies such as Sony, HTC, and Microsoft have also created new platforms for developers to create applications and transmit information from as opposed to the commonly used monitor,  keyboard, and mouse in conjunction with a desktop computer. Interactions utilizing these applications within virtual environments can take advantage of "any perceptual channel, including visual, auditory, haptic, or olfactory," (Bailenson et al., 2008, p. 103-104) as well as "use natural means of communication . . . with immersive VR" (Kauffman, 2003, p. 1) in order to easily pick up and interact with virtual elements on a digital display with a minimal learning curve.

## Uses of the Technology

### Within Entertainment

Over the past two decades, there has been a general shift towards deeply engaging data accumulation as compared to traditional information consumption such as a monitor set on a desk. This movement, largely influenced by the "research and development being conducted in the [video gaming] community" (Zyda, 2005, p. 25), has spurred growth of headsets both in terms of hardware capabilities as well as marketing applicable software to various demographics such as artists and video gamers. Though Nintendo's Virtual Boy was largely considered a commercial failure soon after its 1995 release as according to the Virtual Reality Society, developers from other companies were able to use consumer criticism to better improve their other products for future projects (2016, para. 19). Current technologies that utilize audio and spacial, immersive imagery tend to convey both an intense sense of realism to the user as well as an influx of analytic information to game and graphic designers, allowing them to better improve their platforms as a whole (Zyda, 2005, p. 29). Haptic engine development by use of vibrating controllers additionally allows developers to gradually improve the quality of "sensory stimulation", providing the user with emulated effects upon performing an action. Usage of sensory haptic feedback to simulate recoil from virtual firearms or pinpointed tremors to simulate a dice roll (Zyda, 2005, p. 30) better immerse their players within their virtual environments. Virtual reality developers have also taken advantage of online distributors such as Valve's Steam platform to sell games to consumers, who can write reviews of their products and give constructive feedback to improve products. Many pieces of software additionally include local and online cross-platform multiplayer support, allowing numerous users with Oculus Rift, HTC Vive, or Windows Mixed Reality headsets to use a given application simultaneously in the same virtual space no matter the real-world distance between them. The popularization of the simplistic precision of rhythm games such as Beat Games' *Beat Saber* (2018) or Drool's *Thumper* (2016) contrast massive open-world adventure games such as Besthesda Softworks' *The Elder Scrolls V: Skyrim VR* (2018) or Frontier Developments' *Elite Dangerous* (2015), demonstrating the sheer, spacious spectrum of creativity brought by software developers to consumers around the world. While competition between independent developers and triple-A titles drive consumer markets to buy both hardware and software, a social stigma has been born that these titles are the only reason one should purchase a virtual reality headset. In fact, a straightforward Google search for "VR headset" presents magazines such as PCMag, Digital Trends, and Tom's Guide, all enticing users to join "the next dimension in gaming" (Greenwald, 2018) in their respective lists entitled some variation of "The best VR headsets of 2018" (Wagner, 2018), firstly highlighting each head-mounted display's "growing library of games" (Smith, 2018) before most anything else short of the name of the device itself, including specifications, alternate features, or ultimate cost of the hardware.

**Within Education**

Alternatively, virtual reality technologies have been utilized in job training as well as classroom settings in order to educate both students and professionals to perform tasks. Against the precedent that immersive technologies should solely be promoted to use with video gaming established by media companies to reach the largest audience, virtual reality has been utilized by professionals in a range of fields to improve knowledge of their clients and their workspace as a whole.

**Surgical procedures and medicine.**

An argument made by Grantcharov et al. (2003) cites virtual reality simulators as greatly beneficial “as means of training and objective assessment of psychomotor performance” due to their ability to “allow repeated practice of standardized tasks and provide unbiased and objective measurements of . . . performance”, as evidenced in their experiment of doctors undergoing “education in minimally invasive surgery” procedures such as laparoscopy (p. 146). The results of their 2003 study demonstrated a “significantly greater improvement in performance in the operating room” (p. 146) between those who underwent virtual reality simulator training and those who did not in their control group. By allowing for the creation of controlled environments, educators are able to teach others skills and assess a student’s capabilities and relative knowledge in real time (Rizzo and Kim, 2005, p. 121). Without fear of outside influence, these controlled, virtual environments can be infinitely modified in a precise, managed manner “under a range of stimulus conditions [that are] not easily deliverable and controlled in the real world” (p.119). By eliminating the need to account for errors due to naturally influenced factors such as relative humidity or air pressure during experiments, which are more difficult to precisely and reliably pinpoint and maintain without specific equipment, scientists do not need to worry about altered results due to unreliable or unconsidered variables.

**Rehabilitation and physical therapy.**

Virtual reality technologies are applicable within rehabilitation, clinical trials, and therapy as well, such with Rothbaum et al.’s (1995) study in which college students diagnosed with acrophobia underwent “virtual reality graded exposure” (p. 628) therapy. Proving the hypothesis that immersing oneself within an experience can lead to therapeutic treatment, students who underwent the experiment reported greatly reduced levels of anxiety and improved attitudes towards heights compared to those who did not undergo the simulated assessment (p. 627-628). Grantcharov et al.’s (2003) and Seymour et al.’s (2002) clinical trials concluded similar results, demonstrating exceptionally greater operating room performance from those who underwent virtual reality training compared to those who did not. In Grantcharov et al’s 2003 study, those who “received virtual reality training [not only] performed a laparoscopic cholecystectomy significantly faster than the control group” (p. 146), but also made fewer errors and unnecessary movements; likewise, the “ST group” in Seymour et al.’s 2002 study committed approximately “six times as many errors as the VR group” (p. 461), further validating the claim that virtual reality technologies are viable training tools.

Additionally, Seymour et al. (2002) noticed that demonstrations utilizing simulated, digital elements and virtual environments allowed test supervisors to teach participants important skillsets that would otherwise be impossible to do so without physically existing in a real scenario in an operating room (p. 463). Moreover, a practice experiment in a physical laboratory was not found to be practical due to the fact that “[Operating room] time is so precious and expensive that it can’t actually be used for practice” (p. 464), whereas virtual reality simulations enter as a more effective alternative in cost as well as overall outcome. Combined with the ability to teach complex subjects in a concentrated manner, testing subjects also profited from practicing as many times as necessary in order to fully comprehend material on an individual basis, thoroughly improving the quality of knowledge acquired by the participants while reinforcing the principle of learning by repetition (p. 463). Seymour et al. (2002) concluded that not only can students be exposed to virtual training earlier in their lives, but they can also provide access to both training as well as real-time assessments of student performance. By recording responses of student activities, researchers are able to acutely and objectively judge the performance of participants “at every step of the training phase” (p. 463), determining their readiness for a given topic at a moment’s notice.

Observations gathered from Rizzo and Kim (2005) also cite an expansion of the adoption of virtual reality technologies within the field of medical physical therapy, noticing an increase in material engagement among those children with autism or attention deficit hyperactivity disorder (p. 119, 127). Studies from the American Burn Association regarding post-burn passive range-of-motion physical therapy demonstrate that virtual reality technologies have “[provided] a clinically meaningful degree of pain relief to burn patients” when undergoing standard analgesic care with and without a virtual distraction (Sharar et al., 2007, p. S43). The study found patients experiencing VR as a distraction were much more able to cope with the pain of their injury while being more “cooperative with and more consistent in their therapy, potentially enhancing rehabilitation success” (p. S47) in the long run. Over the course of 3 studies, Sharar et al. reported a 20% reduction in “worst pain intensity”, a 26% reduction in “pain unpleasantness”, and a 37% reduction in “thinking about [their] pain”, regardless of “age, sex, ethnicity, size of initial burn injury, or duration of therapy session” when using virtual reality as an immersive tool for distraction (p. S43). Even a simple interactive environment such as SnowWorld running on Windows 2000 hardware was able to give users a sense of realism and “fun”, demonstrating that technologies can be utilized to benefit others even if they do not have the proper resources for an expensive head-mounted display, as although the experiment was likely very expensive when it occurred, the progression of hardware today would allow software developers to create a similar exposition on a device much less costly (p. S45). Additionally, graphics capability between the era of SnowWorld and today has greatly increased, allowing consumers to deepen their personal connection with what is being displayed on-screen as opposed to low-resolution textures with polygonal, sharp-edged objects.

**Virtual Environments**

Although archaic virtual environments such as SnowWorld (Sharar et al., 2007) have been determined to be effective since the early 2000's, both Carrozzino and Bergamasco (2010) as well as Bailenson et al. (2008) mention the recent increase in use, popularity, and viability of virtual environments (VEs) associated with the enhanced learning experiences within such context. An important subset of virtual reality technologies (as Carrozzino and Bergamasco (2010) define the term "virtual environment" itself as "a complex technology which exploits . . . computer science . . . in order to create a digital environment which users feel completely immersed inside, and which they may interact with" (p. 453)) that emphasizes the value of the user feeling as if they are physically present within a given space through use of haptic feedback and/or visual cues. Such a feeling of "presence" has ultimately become the most prevalent indication of a successfully functioning piece of software in the eyes of many outstanding developers. By transforming how people connect with one another, as indicated by results of Bailenson et al.'s (2008) investigation on the constructs of behavioral and contextual social interaction, virtual environments can "enable transformed social interaction (TSI)" as well as "[uniquely] alter" learning spaces by allowing "teachers and students to use digital technology to strategically alter their online representations and contexts" (p. 102-103) in the near future. For example, the four experiments in Bailenson et al.'s (2008) study demonstrated that virtual classroom environments in which teachers had "augmented social perception" (p. 102) (a simulated environment in which a "teacher either did or did not receive real-time information" (p. 112) on how long they look at each student in the given test room) and students could "[break] the rules of spatial proximity that exist in physical space" (p. 103) by virtually existing closer to the teacher and centered in their field of view (FOV) demonstrated a greater level of engagement between teachers and students as well as an overall greater level of attention to the information being taught to classroom students. (p. 102-103, p. 112-113). A study by Hannes Kaufmann (2003) additionally counters the notion that the insertion of VR technologies would adversely affect overall senses of collaboration and social interaction traditionally brought with a classroom setting by specifically noting the preservation of the ability for teachers to collaborate with students by "mix[ing in] immersive VR or remote collaboration," (p. 1), for those experiencing a lesson in the classroom or even for those unable to attend a given class.

Bailenson et al. (2008) additionally outlined various advantages of utilizing virtual environments as learning tools: Virtual teaching assistants that can specifically produce a multitude of outcomes based on an infinite quantity of variables are one such advantage, allowing them to deliver "personalized one-on-one learning experiences tailored to the individual" (p. 106-107) which would otherwise be either impossible to entertain or financially unreasonable. Further, although VEs inhibit physical, social interaction within a classroom space, high-performing virtual cooperative learning partners can alternatively be substituted in to work with a user and have been found to enhance that student's own performance (p. 107-108). As long as a simulated environment can trick the user into believing they are physically present in the VE itself, the student can uniquely pick up information and mannerisms based on the way they are being exposed to the information, whether it be through assuming the role of a fictional character, a fellow student, or themselves (p. 109). Complex maps of information derived from "multiple perspectives of the same scenario" (p. 108) can also allow users to synthesize information and acquire new insight on the material being taught, such with making more educated judgements on difficult problems based on the information provided to them.

Overall, studies that outline optimal practices of virtual environments as well as practical uses of such technologies allow future software developers to prioritize creating applications that would be most effective and most widely utilized within real-world scenarios.

## Speech and Oral Placement Therapy

Although speech therapy as a field pertains to the assistance of people with language difficulties including cognitive and expressive impairments (Crenshaw 2015), oral placement therapy (OPT) in particular "is a speech therapy which utilizes a combination of [auditory, visual,] and tactile stimulation to the mouth to improve speech clarity" (TalkTools 2018b). Patients undergoing oral placement therapy range in age and speech ability levels and have a varying set of speech disorders (TalkTools 2018b). Individuals treated with oral placement therapy typically include those with autism, cerebral palsy, and Down syndrome as evidenced by similarly associated organizations' affiliation with oral placement tool company TalkTools (2018a).

## Oral Movement and Behavior

Exercises with oral muscular structures allow better mouth and facial movement and behaviors, enabling better speech comprehension between the sender and receiver of a sentence. Along with linguistic oromotor speech patterns, such as “vocalizations . . . , babbling . . . , and speaking” (p. 2), facial movements and “non-speech behaviors such as sucking and chewing” (p. 2), as found by Wilson, Green, and Moore (2008), may appear as soon as the first year of a child’s life. Wilson, Green, and Moore also demonstrate that muscular and jaw movement can be traced in developing fetuses via ultrasound as early as 10-14 weeks (p. 2).

Although specific neural models on human activity pertaining to changes in sucking motions are relatively unknown, models of animal behavior are “relatively well understood” (p. 4) when monitoring changes in adaptation over the first few years of life. For example, nursing children have to learn to inertly adapt their behavior based on outside factors and influences, such with “variations in nipple size and shape[, such with comparing nursing and using a bottle]” (p. 4) as well as “changes in bolus volume, flow, and consistency” (p.4) of fluid entering their bodies (Wilson, Green, and Moore, 2008). However, some children are simply unable to adapt to these changes in environment, and must be assisted to fully conform to their surroundings. For example, underdevelopment in children with Down syndrome causes struggling from an overall inability to effectively manipulate and swallow food on their own, and lack “the motor coordination necessary for normal feeding” until later in life (Hennequin, Faulks, Veyrune, and Bourdiol, 1999, p. 276). These patients have also been shown to have an irregular palette, presenting them with discomfort and an inability to find a natural resting position for their tongue (p. 275-276). Those with Down syndrome thus advance their lower jaw, causing their tongue to protrude out of their mouths and may develop “clenching or grinding [habits in order to] find a position of comfort” (p. 276). Coupled with functional complications regarding postponed “dental eruption” and “behavioral problems such as the refusal to swallow [or] spitting out of food” (p. 276), many find it difficult to care for and live as Down syndrome patients, ultimately affecting their demeanor, health, and livelihood when participating in activities typically considered normal for any given control subject.

## Conclusion

Ultimately, as described previous studies, the use of virtual reality technologies should be further experimented upon in order to definitively determine the practical usage of a given training scenario for oral speech therapists. As proven by a variety of applications being created by VR software companies around the world, even subfields of medicine, mechanics, and psychology would greatly benefit from the addition of virtual experiences brought with new hardware; however, there is currently only a minuscule percentage of businesses that do, in fact, make full use of the technology compared to those who do not. More specifically on the overarching topic of oral speech therapists, if fundamentally adapted to its optimal capacity, therapists will be able to more effectively use devices to assist children and adults with optimal muscular movement when speaking and resting naturally. Not only will therapists be able to more effectively use tools, however, but they will be able to pinpoint exact muscles to work with and solve both physical and cosmetic problems that some patients with Down syndrome or autism face.

## Methods to Determine the Ability of Virtual Reality Technologies as Pertaining to Speech Therapy

The following methodological approach described deduces the applicability of virtual reality systems in the context of educational learning, specifically within oral placement therapy device training. The researcher used deductive reasoning to determine whether virtual reality technologies can be effectively utilized in order to optimally train oral placement therapists by distributing surveys containing qualitative and quantitative inquiries. Surveys compared a control group of participants viewing a traditional lecture in 2D of an overview of Oral Placement Therapy with a test group of participants utilizing virtual reality headsets to view a video of identical subject matter in 3D. Results were analyzed and contrasted to determine answers to questions and subquestions pertaining to if virtual reality technologies can be utilized in order to train oral placement therapists to improve comprehension, usage, and effectiveness of speech therapy tools.

## Participants/Subjects

All participants of the study were clients of TalkTools, a local designer and manufacturer for oral and speech therapy devices. All clients were also professional therapists who have prior experience with TalkTools products. Responses themselves were collected anonymously to protect the identity of participants. Instructional materials were delivered via email, which included a link to a Dropbox folder containing both 2D and 3D videos, an instructional ReadMe file containing a link to the Google Form, and a thank-you message for participating in the study. Additionally, those selected to use the 3D video were shipped generic plastic VR viewers from amazon.com to use with their own smartphones. Involvement in the study after being contacted to participate was voluntary, but those randomly selected to test virtual reality video were allowed to keep the plastic virtual reality viewer if so desired.

## Apparatus/Measures/Materials

Ware and Franck (1994) found that the amount of "information [that] can be perceived in [a 3D diagram is much greater than its] 2D [counterpart]" (p. 182), which was supported by Liarokapis et al.'s 2004 study where researchers found an increase in comprehension of the operation of mechanical systems after viewing digitally simulated three-dimensional web content. These claims support the idea that 3D video should be used in order to increase comprehension of teaching material. Equipment utilized within this project included the use of an Oculus Rift head-mounted display connected to a desktop computer to develop and modify stereoscopic 3D video. The researcher utilized scaleable, virtual models of anatomically accurate internal structures taken from Medis Media's 3D Organon VR, an application found on the Oculus Store. The application being run within the headset was then mirrored to the desktop monitor using the Oculus Mirror diagnostic tool run from Windows command prompt with modifier --PostDistortion (Medis Media, 2018). A path to the Oculus Mirror executable as well as command line modifiers can be found at https://developer.oculus.com/documentation/pcsdk/latest/concepts/dg-compositor-mirror/ (Oculus VR, LLC, 2018a). The researcher recorded video of the application using the built-in Windows 10 Game Bar utility to correspond with audio from a previously-analyzed TalkTools lecture. Timestamps of each structure were handwritten to aid the recording process. The researcher then compiled and edited the final videos for distribution using the macOS software Final Cut Pro X v10.4.2.

## Procedures[2]

The researcher firstly proceeded with data collection by formatting and organizing both stereoscopic three-dimensional and traditional two-dimensional video into an organized lesson with set parameters, audio, and instructional content to garner responses. Timestamps were recorded externally on paper to reference while recording and editing the video. Plastic virtual reality viewers were then distributed to clients who had been selected for participation in the experimental test group, and all participants were sent a link to a shared Dropbox folder containing instructional videos to load onto their smartphones. Questions were compiled by the researcher and reviewed by professionals not participating in the study, and data was collected and analyzed via Google Form, comparing and contrasting client responses and reporting on common themes, frequently chosen responses, and general opinions of those participating.

This study has collected both quantitative and qualitative measures. Participants were both graded on information retained after viewing the video, which was appropriately named “Section 1: Knowledge Test”, as well as personal reception to the video itself, which was similarly named “Section 2: Opinion Test”. In Section 1: Knowledge Test, participants were quantitatively quizzed using a series of 6 multiple choice questions, each worth 5 points, and were notified with their score out of 30 points after they completed the entire test. Section 2: Opinion Test used both quantitative and qualitative measures to garner responses, with quantitative questions utilizing a Likert scale ranging from 1-5, with 1 being “Strongly Disagree” and 5 being “Strongly Agree”. Other qualitative questions were used for diagnostic purposes to determine whether the participant viewed the 2D or 3D video and receive feedback regarding how they believe a given presentation should be improved to increase the retention of information. All questions were manually graded and analyzed through Google Forms and sorted in a spreadsheet.

[2] A complete flowchart explicitly outlining all steps taken to produce the modified, 3D video used in this study can be found in Appendix C on page 49.

## Evaluation

The data collected is proven valid as a large majority of responses based on a large, randomized subset found a correlation with one another and allowed a conclusion to be drawn; however, due to the fact that the video used in this study was simply immersive and not interactive, some users found some faults in the video's construction itself, demonstrating that further research is required using more advanced software. The data has determined that some of those experiencing a different version of the same information enjoy and/or prefer use of immersive technology as opposed to traditional lecture, demonstrating that virtual reality is viable as a learning tool in the oral placement field, according to TalkTools clients, as long as technology is improved.

## Conclusion

Conclusions supported by the data has confirmed the hypothesis that virtual reality technologies are effective as immersive learning tools within the subset of oral placement therapy. While the results of this study are not entirely all-inclusive due to the small subject pool, specificity of the field, and varying needs and preferences of individual subjects, the aim of the study as a whole was to broaden the reach of and bring awareness to the usage of virtual reality technologies in fields apart from the commonly assumed areas of media and entertainment, which has been accomplished.

## Results Pertaining to the Ability of Virtual Reality Technologies to Improve Comprehension of Speech Therapy Device Training

Based on results to the survey including optional feedback provided from participants, this chapter includes significant findings of the research described previously. Data was collected from oral placement therapists who viewed instructional material via Google Form[3] that was then manually evaluated.

## Results

### Section 1: Knowledge Test

Section 1, the fact recall portion of the survey, contained six (6) questions gauging participant recollection of data from their informational video. Participants were provided with three (3) multiple-choice questions and three (3) true/false questions to answer. Participants were originally given five (5) points for every question answered correctly and zero (0) points for every question answered incorrectly for a sum total of thirty (30) points; however, due to the fact that the first fifteen responses to the second true/false question "[T/F] Air is shaped by the LIPS, VELUM, and TONGUE to articulate speech" were all answered incorrectly, the researcher ultimately decided to omit the question (only 29.17% of participants answered the question correctly overall ("FALSE")). Thus, although each participant noted their scores out of 30 after taking the survey, the end results of the experiment do not factor in that question for a grand total of 25 points instead. Each question was written using information directly addressed in the video participants viewed, although some contained more specific inquiries than others. 19 professional therapists from the experimental "3D" group and 5 professional therapists from the control “2D” group participated in the survey. Since scores were calculated out of 25 points instead of the original 30, the average score from all participants was 80.80%; however, each group received their own range of scores, with 3D receiving anywhere from 10-25 points (with an average score of 77.80%), while 2D’s was much more condensed from only 20-25 points (with an average score of 92.00%). All participants answered at least 2 multiple choice questions correctly.

[3] A full copy of the Google Form utilized in this study can be found in Appendix A, starting on page 39.

**Section 2: Opinion Test**

Section 2, the personal opinion portion of the survey, contained six (6) questions gauging participant reactions to the data presented in the video as well as how the data was presented to them. As the survey itself was conducted anonymously, participants were firstly asked which study group they tested for, whether that be the “2D” or “3D” group, for diagnostic purposes to filter their results. Subsequently, the form asked participants to rank their agreement of 4 given statements using a Likert scale, ranging from 1 (“Strongly Disagree”) to 5 (“Strongly Agree”). Additionally, participants in the 3D group were asked whether or not they preferred that method of learning as compared to a traditional lecture (2D participants were instead asked to select the “I did not view the presentation using a headset” option. Finally, all participants were provided with the opportunity to give a personal statement with feedback regarding how they would personally better the teaching of material to themselves. Of the 4 Likert scale cues, the researcher considered the first 3 “positive” in that they elicit quantifiably higher responses for more favorable feedback. Conversely, the researcher considered the final prompt, “The presentation gave me motion sickness”, “negative”, in that it elicited a quantifiably lower response for favorable feedback, as optimally a presentation should not typically desire to induce sickness in a participant. Overall, those viewing the 2D video seemed to have an overall higher reception in terms of enjoyment, knowledge gained, and engagement on average; however, there was a greater quantity of higher responses in the 3D group. In fact, there are five “optimal” responses recorded in the 3D group, answering 5s for each “positive” prompt and a 1 in the “negative” prompt, while only one such was recorded from those who viewed the 2D content. Responses illustrated that personal opinions about the content were largely polarized; participants generally either really enjoyed or hardly at all enjoyed the presentation, with few average responses. The responses to the prompt for those only in the experimental group demonstrates the polarization previously described, as there lies a 50-50 split in the data between those who would prefer virtual reality, 3D immersive lessons over traditional lectures and those who would not prefer such. Optional feedback generally echoed numerical responses provided by the given participant, although interesting replies included the lack of subtitles for reference as well as one approval of the instructional material itself in Spanish. Otherwise, written responses from the 3D group gave examples as to their grievances with the hardware, generally due to how the video was constructed as well as the hardware used as opposed to the learning method itself.

**Trends**

Although the sample size is significantly (73.69%) smaller, the 2D group scored a decent quantity (14.20%) higher overall on the knowledge test portion of the survey. Participants in the control (2D) group also scored 23/25 points (92.00%) on average, compared to the experimental (3D) group’s score of 19.47/25 points (77.88%). Since a score of 3 was considered a “completely average” score in this experiment, as a Likert scale from 1 to 5 would render an average score at 3, both tests received favorable positive (average > 3) and negative (average <

3) ratings. The 2D group saw an average score of 4.53/5 over positive responses and 1.8/5 over the negative response, and the 3D group saw an average of 3.82/5 and 2.63/5. Exactly half of 3D experimental participants (8/16) preferred such a learning method over a traditional lecture (3 participants (15.79%) selected “No preference” instead). However, due to the great disparity in the quantity of control group participants, the data presented in this section is not the greatest comparison.

## Question Analysis

Results gathered in this study seek to answer the following question: “How can virtual reality technologies be utilized in order to optimally train oral placement therapists and improve the comprehension, usage, and effectiveness of speech therapy tools?” and have done so by investigating one possible change that could be made to enhance the quality of oral placement therapy tool use. Analyzed feedback from professionals in the field has noted that some would prefer further lessons on using tools in virtual reality; however, the hardware itself would need to be modified for improved ease-of-use, and the video would need to be slightly slowed to improve comprehension of material being taught. The quality of the video, downscaled to 720p, may also need to be slightly improved in order to assist participants with discerning information presented to them on-screen. Additionally, as stated in the review of literature pertaining to virtual reality technologies and speech therapy device training, virtual reality technologies have transformed from art to entertainment to medicine, undergoing numerous alterations to improve one’s sense of immersion in a given landscape. Improvements of virtual environments have assisted with heightening one’s sense of immersion through the addition of creative assets and perspectives to imagine an infinite quantity of digital worlds to immerse oneself into. Such assets, becoming more detailed as the technology itself improves, effect the visualizations of content by allowing a user to study features at an unparalleled scope as precise and otherwise minute aspects of a given item, such as muscular systems, can be resized to larger proportions. Ultimately, although not all users prefer to use virtual reality technologies when they could instead simply view a lecture, VR has been found to enhance how people can visualize their own work.

## Discussion of Results Pertaining to the Ability of Virtual Reality to Improve Comprehension of Speech Therapy Device Training

This chapter analyzes the data presented in the previous section including contributions to the field. Conclusions drawn on the information as a whole can also be found.

### Overview of Content; Feedback

Although each participant only viewed 3 minutes and 18 seconds worth of content, most all therapists were able to answer at least 60% of the questions correctly. The optional final question on the survey, “Optional: What improvements would you make to the presentation to better convey the information being taught?” received 14 responses out of the total 24 participants, 13 of which were in English and 1 of which was in Spanish. Only one (1) response was recorded from the 2D group, and the remaining 13 were from the 3D experimental group. Overall, response feedback varied across the board; however, trends emerged among those who previously stated that they did not prefer the 3D video from 2D, typically being that actually viewing the content, not the presentation method of the content itself, heavily influenced their opinions. Unfortunately, since the researcher cannot physically ensure that each participant has a phone that is large enough to render the video in high quality while retaining a large screen size, then results are subject to slight bias. For example, one participant noted that, due to the fact that they used an iPhone 5 (screen size: 4” diagonally), they could not focus the 3D video, rating it a 5 for the “The presentation gave me motion sickness.” response after viewing “2 videos of Sara side by side”. It should be noted that videos were tested using an iPhone 7 Plus (screen size: 5.5” diagonally) and a Samsung Galaxy S8 (screen size: 5.8” diagonally), which are much larger than an iPhone 5. Additionally, the speed of the video itself was a little too fast for therapists to adequately comprehend the information being taught. Combined with a lack of subtitles, many found it difficult to comfortably focus on and follow along with the video, especially for those "with English as a second language", which the researcher failed to consider when creating demonstration videos. Ultimately, these problems could be eliminated by standardizing the device used to watch the supplied material and developing a specialized application to more carefully detail only necessary information to a given lesson.

## Contributions and Resolutions

Contributions to the field of virtual reality development as it pertains to oral placement therapy are currently largely inconclusive due to the fact that responses largely narrowed down to personal factors such as whether or not the user could adequately view the video or simply did not enjoy viewing it; however, based on the data currently gathered and feedback received, VR can be more effectively utilized as long as the video created is well-executed and all participants have an easy method of viewing the video with a larger smartphone. This study has helped resolve the original research question "How can virtual reality technologies be utilized in order to optimally train oral placement therapists and improve the comprehension, usage, and effectiveness of speech therapy tools?" through the following methods:

- Virtual reality technologies allow users to more effectively see *within* their patients on a larger, more immersive scale.
- Virtual reality technologies can be utilized to train oral placement therapists in video format, although it might not be the most effective method.
- Can be used in video format with text and subtitles, but they must be slowed down to allow users to focus on and read individual names of muscles or structures.

- True adoption on a personal level is ultimately determined by personal preference as to if a given user enjoys using the headset or not.

## Future Research

Future research should ensure a constant device for all testing purposes, such as a singular device with large screen size (i.e. all Samsung Galaxy S8 or iPhone 7+ devices, as tested by the researcher), preferably one that can render a program instead of a video to allow users to learn material at their own pace, apart from a video where their view becomes static. Allowing each participant to utilize the same or an identical Oculus Rift as utilized by the researcher may also solve this issue and greatly improve reactions to the learning method. Additionally, in order to minimize preconceived bias, prior individual user history with virtual reality should be considered in order to find users who have not been exposed to or have the same level of exposure to VR technology. Researchers conducting future study should also ensure that participants selected have no previous exposure to material being taught in order to garner an equal comparison. Optimally, each participant would be able to explore with their own entire model to navigate instead of a video; however, until hardware becomes massively available and reasonably affordable, that is unlikely to occur on a wide-scale as many would not see a rational use in adopting the technology at such a time, even if it has been proven to increase comprehension of information.

# Appendix A

## Post-Video Survey

The attached survey on the following page (40) was exported directly from Google Forms on October 10, 2018 (https://goo.gl/NyMoS6: “Post-Video Survey”). No modifications were made to the form after data collection began.

# Post-Video Survey

Please answer this Google Form after you have viewed the short video attached in the Dropbox folder. This survey is completely anonymous and should take less than 10 minutes to complete.

* Required

## Section 1: Knowledge Test

This section reviews information covered in the video.

1. **Can Oral Placement Therapy be used interchangeably with speech therapy?** *
   *Mark only one oval.*
   - Yes
   - No

2. **Where does speech movement begin?** *
   *Mark only one oval.*
   - Mouth
   - Back of the throat
   - Velum
   - Abdomen

3. **If the velum moves AWAY from the pharyngeal wall, where does airflow go?**
   *Mark only one oval.*
   - Mouth
   - Nose
   - Tongue
   - Abdomen

4. **[T/F] The jaw is the support system for the independent movement of the lips and the tongue for BOTH feeding and speech.** *
   *Mark only one oval.*
   - True
   - False

5. **[T/F] Air is shaped by the LIPS, VELUM, and TONGUE to articulate speech.** *
*Mark only one oval.*
   - True
   - False

6. **[T/F] Speech clarity is one, wholistic muscle movement.** *
*Mark only one oval.*
   - True
   - False

## Section 2: Opinion test

This section reviews how the viewer received information using a Likert scale. Please answer each question completely honestly.

7. **How did you view the video?** *
*Mark only one oval.*
   - In 2D, with my computer or mobile device
   - In 3D, with a mobile device and a headset

8. **I enjoyed the presentation.** *
*Mark only one oval.*

| | 1 | 2 | 3 | 4 | 5 | |
|---|---|---|---|---|---|---|
| Strongly Disagree | | | | | | Strongly Agree |

9. **I feel like I learned something from the presentation.** *
*Mark only one oval.*

| | 1 | 2 | 3 | 4 | 5 | |
|---|---|---|---|---|---|---|
| Strongly Disagree | | | | | | Strongly Agree |

10. **I feel like I was engaged with the material being presented.** *
*Mark only one oval.*

| | 1 | 2 | 3 | 4 | 5 | |
|---|---|---|---|---|---|---|
| Strongly Disagree | | | | | | Strongly Agree |

11. **The presentation gave me motion sickness.** *
*Mark only one oval.*

| | 1 | 2 | 3 | 4 | 5 | |
|---|---|---|---|---|---|---|
| Strongly Disagree | ◯ | ◯ | ◯ | ◯ | ◯ | Strongly Agree |

12. **If you viewed the presentation using a headset in 3D, did you prefer this method of learning as compared to a traditional lecture?** *
*Mark only one oval.*

◯ Yes
◯ No
◯ No preference
◯ I did not view the presentation using a headset.

13. **Optional: What improvements would you make to the presentation to better convey the information being taught?**

Powered by Google Forms

# Appendix B

## Raw Data Extracted from Google Sheets

The spreadsheet on the following five pages (44 through 48) detail all 24 raw responses exported from the Google Form, color coded for ease of evaluation.

| | A | B | C | D | E | F | G | H | I | J | K | L | M |
|---|---|---|---|---|---|---|---|---|---|---|---|---|---|
| 1 | Timestamp | Score | Can Oral P | Where doe | If the velum | [T/F] The ja | [T/F] Air is | [T/F] Speec | How di | I enjoye | I feel lik | I feel lik | The pre |
| 2 | 7/13/2018 19:12:44 | 25 / 30 | No | Abdomen | Nose | TRUE | TRUE | FALSE | In 3D, | 5 | 4 | 5 | 1 |
| 3 | 8/28/2018 16:29:29 | 10 / 30 | Yes | Velum | Abdomen | TRUE | TRUE | FALSE | In 3D, | 5 | 5 | 5 | 5 |
| 4 | 8/30/2018 12:31:33 | 10 / 30 | Yes | Velum | Mouth | TRUE | TRUE | FALSE | In 3D, | 1 | 1 | 1 | 5 |
| 5 | 8/30/2018 12:40:18 | 20 / 30 | Yes | Abdomen | Nose | TRUE | TRUE | FALSE | In 3D, | 5 | 5 | 5 | 1 |
| 6 | 8/30/2018 13:43:14 | 10 / 30 | Yes | Velum | Abdomen | TRUE | TRUE | FALSE | In 3D, | 3 | 3 | 3 | 3 |
| 7 | 8/30/2018 14:28:55 | 15 / 30 | No | Velum | Nose | TRUE | TRUE | TRUE | In 3D, | 4 | 4 | 3 | 1 |
| 8 | 9/10/2018 20:17:11 | 25 / 30 | No | Abdomen | Nose | TRUE | TRUE | FALSE | In 3D, | 4 | 3 | 5 | 4 |
| 9 | 9/11/2018 15:21:53 | 20 / 30 | Yes | Abdomen | Nose | TRUE | TRUE | FALSE | In 3D, | 2 | 3 | 1 | 4 |
| 10 | 9/13/2018 12:29:13 | 15 / 30 | Yes | Abdomen | Mouth | TRUE | TRUE | FALSE | In 3D, | 1 | 1 | 1 | 5 |
| 11 | 9/13/2018 13:06:39 | 15 / 30 | Yes | Abdomen | Nose | TRUE | TRUE | TRUE | In 3D, | 5 | 5 | 5 | 1 |
| 12 | 9/20/2018 18:30:39 | 25 / 30 | No | Abdomen | Nose | TRUE | TRUE | FALSE | In 3D, | 5 | 5 | 5 | 1 |
| 13 | 9/24/2018 14:01:29 | 20 / 30 | No | Abdomen | Nose | TRUE | TRUE | TRUE | In 3D, | 5 | 5 | 5 | 1 |
| 14 | 9/27/2018 17:24:46 | 20 / 30 | No | Abdomen | Nose | TRUE | TRUE | TRUE | In 3D, | 5 | 5 | 5 | 4 |
| 15 | 10/4/2018 17:30:16 | 15 / 30 | Yes | Abdomen | Nose | FALSE | TRUE | FALSE | In 3D, | 5 | 5 | 5 | 3 |
| 16 | 10/8/2018 23:18:32 | 25 / 30 | No | Abdomen | Nose | TRUE | TRUE | FALSE | In 3D, | 3 | 4 | 3 | 5 |
| 17 | 9/27/2018 16:54:23 | 30 / 30 | No | Abdomen | Nose | TRUE | FALSE | FALSE | In 2D, | 5 | 4 | 4 | 1 |
| 18 | 10/4/2018 16:56:41 | 25 / 30 | No | Abdomen | Nose | TRUE | FALSE | TRUE | In 2D, | 5 | 5 | 5 | 3 |
| 19 | 10/26/2018 16:15:25 | 30 / 30 | No | Abdomen | Nose | TRUE | FALSE | FALSE | In 2D, | 5 | 5 | 5 | 1 |
| 20 | 10/27/2018 2:52:34 | 25 / 30 | No | Abdomen | Nose | TRUE | TRUE | FALSE | In 2D, | 5 | 5 | 4 | 3 |
| 21 | 10/28/2018 6:51:23 | 20 / 30 | Yes | Abdomen | Nose | TRUE | TRUE | FALSE | In 2D, | 5 | 5 | 5 | 1 |
| 22 | 10/29/2018 13:25:57 | 30 / 30 | No | Abdomen | Nose | TRUE | FALSE | FALSE | In 3D, | 4 | 4 | 4 | 3 |
| 23 | 10/31/2018 11:34:18 | 30 / 30 | No | Abdomen | Nose | TRUE | FALSE | FALSE | In 3D, | 4 | 3 | 3 | 1 |
| 24 | 11/1/2018 9:55:36 | 30 / 30 | No | Abdomen | Nose | TRUE | FALSE | FALSE | In 3D, | 3 | 3 | 2 | 1 |
| 25 | 11/30/2018 14:56:26 | 30 / 30 | No | Abdomen | Nose | TRUE | FALSE | FALSE | In 2D, | 3 | 4 | 4 | 1 |

| N | O | P | Q | R | S |
|---|---|---|---|---|---|
| If you vie | Optional: What improvements would you make to the presentation to better convey the information being taught? | | | | |
| Yes | | | | | |
| Yes | | | | | |
| No | I think my results are skewed due to the fact I have an iphone 5. I was not able to get the video in focus, and, instead, watched 2 vi | | | | |
| Yes | Kind of difficult to use if you wear glasses | | | | |
| No | I liked the headset but didn't like the 3D | | | | |
| No | First, the 3D headset gave me double vision. Second, it did not work well with my phone design. | | | | |
| Yes | I feel it's very cool but WAY too fast you don't need all the muscles labeled it's too hectic . Go slower and label specifically the musc | | | | |
| No | The movements around the body were too quick and not consistent with what was being discussed. I was distracted by trying to re | | | | |
| No | The picture in the video was way to close to my eyes. I wasn't able to focus on the video. If I closed one eye it was better. When the | | | | |
| No | | | | | |
| Yes | Maybe I didn't have the phone positioned correctly but the writing with the muscles was somewhat blurry. | | | | |
| Yes | Closed caption in English for professionals with English as a second language. | | | | |
| No prefe | | | | | |
| Yes | | | | | |
| No | Viewing the anatomy was good, especially from the inside lookingout, but too close to my eyes for comfort. | | | | |
| I did not | | | | | |
| I did not | Todo adecuado. | | | | |
| Yes | It is difficult to comment having seen such a small portion of the presentation. The three demensional pictures were confusing as th | | | | |
| I did not | | | | | |
| I did not | | | | | |
| No prefe | It squished my nose a lot. | | | | |
| No | I thought the graphics used were amazing! I was a bit distracted by them so I wasn't listening as well as I could have. The movemer | | | | |
| No prefe | | | | | |
| I did not | | | | | |

| T | U | V | W | X |
| --- | --- | --- | --- | --- |
| | | | | |
| | | | | |
| | | | | |
| deos of Sara side by side. Headache! | | | | |
| | | | | |
| | | | | |
| | | | | |
| cles targeted and show them so it sticks with the listener . Beautiful idea ! Also mention ebp as Moore said jaw is foundation as van ri | | | | |
| ead the labels of the muscles system, being nauseated when the view was close v. far and couldn't keep up. | | | | |
| e part with the pictures and diagrams came up, I wasn't able to follow what I was looking at. Also the labels in the diagram were goin | | | | |
| | | | | |
| | | | | |
| | | | | |
| | | | | |
| | | | | |
| | | | | |
| | | | | |
| | | | | |
| hey appeared to move rapidly and I would have like more attentiion given to them (labeling execution of movements etc.) | | | | |
| | | | | |
| | | | | |
| | | | | |
| nt of the camera needs to be slower. A way to rewind (without taking the phone out) would be helpful. | | | | |
| | | | | |
| | | | | |

| Y | Z | AA | AB | AC |
|---|---|---|---|---|
| | | | | |
| | | | | |
| | | | | |
| | | | | |
| | | | | |
| | | | | |
| iper said use every tool possible . Consider showing these names and dates somewhere on the screen. | | | | |
| | | | | |
| g so fast I didn't know what to look at. I would definitely not choose to watch a presentation in this manner. Also the elastic strap on t | | | | |
| | | | | |
| | | | | |
| | | | | |
| | | | | |
| | | | | |
| | | | | |
| | | | | |
| | | | | |
| | | | | |
| | | | | |
| | | | | |
| | | | | |
| | | | | |
| | | | | |

| AD | AE | AF |
|---|---|---|
| the top of my head would not stay and I had to hold it the entire time. | | |

# Appendix C

The flowchart on the following page (50) includes a visual aid with step-by-step instructions to reproduce the methods previously outlined.

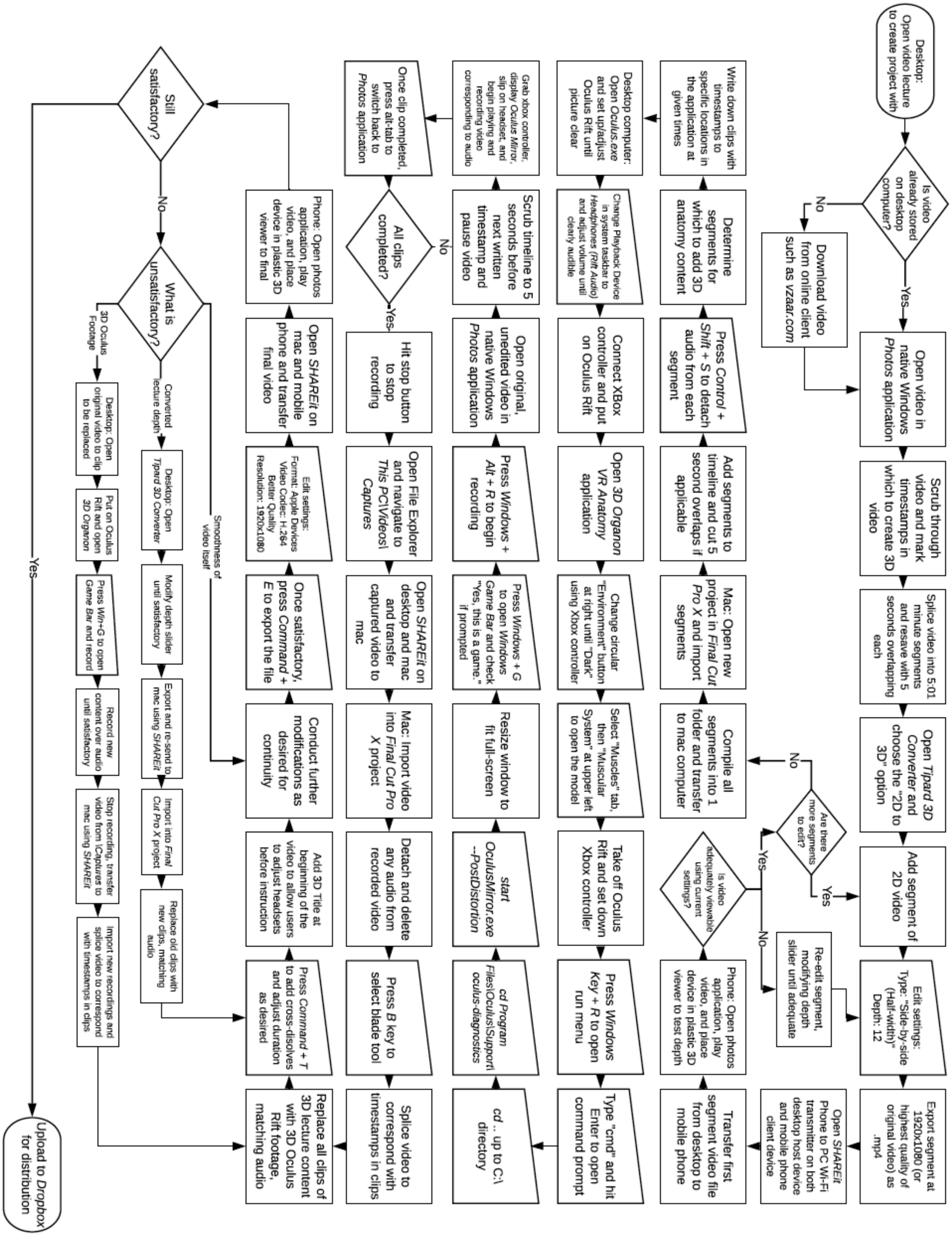
Desktop: Open video lecture to create project with
Is video already stored on desktop computer?
No
Yes
Download video from online client such as vzaar.com
Open video in native Windows Photos application
Scrub through video and mark timestamps in which to create 3D video
Splice video into 5:01 minute segments and resave with 5 seconds overlapping each
Open Tipard 3D Converter and choose the "2D to 3D" option
Add segment of 2D video
Edit settings: Type: "Side-by-side (Half-width)" Depth: 12
Export segment at 1920x1080 (or highest quality of original video) as .mp4
Open SHAREit Phone to PC Wi-Fi transmitter on both desktop host device and mobile phone client device
Transfer first segment video file from desktop to mobile phone
Phone: Open photos application, play video, and place device in plastic 3D viewer to test depth
Is video adequately viewable using current settings?
Re-edit segment, modifying depth slider until adequate
Are there more segments to edit?
Compile all segments into 1 folder and transfer to mac computer
Mac: Open new project in Final Cut Pro X and import segments
Add segments to timeline and cut 5 second overlaps if applicable
Press Control + Shift + S to detach audio from each segment
Determine segments for which to add 3D anatomy content
Write down clips with timestamps to specific locations in the application at given times
Desktop computer: Open Oculus.exe and set up/adjust Oculus Rift until picture clear
Change Playback Device in system taskbar to Headphones (Rift Audio) and adjust volume until clearly audible
Connect XBox controller and put on Oculus Rift
Open 3D Organon VR Anatomy application
Change circular "Environment" button at right until "Dark" using Xbox controller
Select "Muscles" tab, then "Muscular System" at upper left to open the model
Take off Oculus Rift and set down Xbox controller
Press Windows Key + R to open run menu
Type "cmd" and hit Enter to open command prompt
cd .. up to C:\ directory
cd Program Files\Oculus\Support\oculus-diagnostics
start OculusMirror.exe --PostDistortion
Resize window to fit full-screen
Press Windows + G to open Windows Game Bar and check "Yes, this is a game." if prompted
Press Windows + Alt + R to begin recording
Open original, unedited video in native Windows Photos application
Scrub timeline to 5 seconds before next written timestamp and pause video
Grab Xbox controller, display Oculus Mirror, slip on headset, and begin playing and recording video corresponding to audio
Once clip completed, press alt-tab to switch back to Photos application
All clips completed?
Hit stop button to stop recording
Open File Explorer and navigate to This PC\Videos\Captures
Open SHAREit on desktop and mac and transfer captured video to mac
Mac: Import video into Final Cut Pro X project
Detach and delete any audio from recorded video
Press B key to select blade tool
Splice video to correspond with timestamps in clips
Replace all clips of 3D lecture content with 3D Oculus Rift footage, matching audio
Add 3D Title at beginning of the video to allow users to adjust headsets before instruction
Press Command + T to add cross-dissolves and adjust duration as desired
Conduct further modifications as desired for continuity
Once satisfactory, press Command + E to export the file
Edit settings: Format: Apple Devices Video Codec: H.264 Better Quality Resolution: 1920x1080
Open SHAREit on mac and mobile phone and transfer final video
Phone: Open photos application, play video, and place device in plastic 3D viewer to final
Still satisfactory?
What is unsatisfactory?
Smoothness of video itself
Converted lecture depth
3D Oculus Footage
Desktop: Open Tipard 3D Converter
Modify depth slider until satisfactory
Export and re-send to mac using SHAREit
Import into Final Cut Pro X project
Replace old clips with new clips, matching audio
Desktop: Open original video to clip to be replaced
Put on Oculus Rift and open 3D Organon
Press Win+G to open Game Bar and record
Record new content over audio until satisfactory
Stop recording, transfer video from \Captures to mac using SHAREit
Import new recordings and splice video to correspond with timestamps in clips
Upload to Dropbox for distribution